\documentclass[a4paper,11pt,numbers=noenddot]{scrartcl}
\usepackage{libertine}
\usepackage[T1]{fontenc}
\usepackage[utf8]{inputenc}
\usepackage{slantsc}
\usepackage{array}
\usepackage{amsmath}
\usepackage{adjustbox}
\setkomafont{subsection}{\usefont{T1}{fvm}{m}{n}}
\setkomafont{section}{\usefont{T1}{fvs}{b}{n}\Large}
\title{{\LARGE Holding a Conference Online and Live \\ due to Covid-19}}
\subtitle{A Report on EDBT / ICDT 2020 (version: April 16, 2020)}
\author{
  \begin{tabular}{c}
    Angela Bonifati\\[-1mm]
    \small Lyon 1 University\\[-2mm]
    \small (France)
  \end{tabular}
  \and
  \begin{tabular}{c}
    Giovanna Guerrini\\[-1mm]
    \small University of Genova\\[-2mm]
    \small (Italy)
  \end{tabular}
  \and
  \begin{tabular}{c}
    Carsten Lutz \\[-1mm] \small University of Bremen\\[-2mm]
    \small (Germany)
  \end{tabular}
  \and
  \begin{tabular}{c}
    Wim Martens \\[-1mm]
    \small University of Bayreuth\\[-2mm]
    \small (Germany)
  \end{tabular}
  \and
  \begin{tabular}{c}
    Lara Mazilu \\[-1mm]
    \small University of Manchester\\[-2mm]
    \small (UK)
  \end{tabular}
  \and
  \begin{tabular}{c}
    Norman Paton \\[-1mm]
    \small University of Manchester\\[-2mm]
    \small (UK)
  \end{tabular}
  \and
  \begin{tabular}{c}
    Marcos Antonio Vaz Salles \\[-1mm]
    \small University of Copenhagen\\[-2mm]
    \small (Denmark)
  \end{tabular}
  \and
  \begin{tabular}{c}
    Marc H. Scholl \\[-1mm]
    \small University of Konstanz\\[-2mm]
    \small (Germany)
  \end{tabular}
  \and
  \begin{tabular}{c}
    Yongluan Zhou \\[-1mm]
    \small University of Copenhagen\\[-2mm]
    \small (Denmark)
  \end{tabular}
}

\usepackage{enumerate}
\usepackage{xspace}

\usepackage{tikz}
\usetikzlibrary{calc}
\usetikzlibrary{shapes.misc}
\usepackage{pgfplots}
\usepackage{pgfplotstable}
\pgfplotsset{compat=1.8}
\usepgfplotslibrary{statistics}
\usepgfplotslibrary{colorbrewer}

\usepackage{pgf-pie}

\usepackage{url}
\usepackage{hyperref}

\newcolumntype{R}{>{\kern\tabcolsep}r<{\kern\tabcolsep}@{}}

\definecolor{brickred}{rgb}{0.8, 0.25, 0.33}

\newcommand{\eat}[1]{}

\date{}

\begin{document}

\maketitle

\begin{abstract}
  The joint EDBT/ICDT conference (International Conference on
  Extending Data\-base Technology/International Conference on Database
  Theory) is a well established conference series on data management,
  with annual meetings in the second half of March that attract 250
  to 300 delegates. Three weeks before EDBT/ICDT 2020 was planned to
  take place in Copenhagen, the rapidly developing Covid-19 pandemic
  led to the decision to cancel the face-to-face event. In the
  interest of the research community, it was decided to
  move the conference online while trying to preserve as much of the
  real-life experience as possible. As far as we know, we are one of the
  first conferences that moved to a fully synchronous online experience
  due to the COVID-19 outbreak. By fully synchronous, we mean that
  participants jointly listened to presentations, had live Q\&A, and
  attended other live events associated with the conference. In this
  report, we share our decisions, experiences, and lessons
  learned.
\end{abstract}

\section{Introduction}

Three weeks before EDBT/ICDT 2020 was planned to take place in
Copenhagen, the rapidly developing Covid-19 pandemic reached a state
in which it became clear that the face-to-face event had to be
canceled. We, the organizers, then decided to move the conference
online while trying to preserve as much of the real-life experience as
possible. Given the very short notice, we had to be pragmatic and
could not prepare the online event as carefully as we would have done
otherwise. In fact, we considered the whole thing an interesting
experiment suggested by the circumstances, with potentially important
lessons to be learned for the community and beyond. We were delighted
to receive significant support for our decision from the attendees and
keynote speakers and by the community spirit that developed in its
course. The online event ran smoothly and was much more enjoyable and
successful than we had expected. We received a lot of positive
feedback, both informally and in the survey that we sent to our
participants after the conference. The purpose of this document is
to share our experience and the lessons learned, so that other
conference organizers facing a similar situation can benefit from it. 

\section{Decisions}

We go over a list of questions that other conference organizers are
likely to face when considering a move online, share our decisions,
describe how we implemented them in practice, and what we think
about the outcome in retrospect.  Since the conference needed to be
re-organized within a very short time and the effect on internet services of
millions of people working from home because of the pandemic was not
clear, we used an important guiding principle:
\begin{equation}\label{GP}
\text{Aim to minimize our dependency on stable internet connections.}\tag{GP}
\end{equation}

\paragraph{Should the conference be synchronous (i.e., live) or asynchronous?}
We decided to run the event in a fully live mode to simulate a
face-to-face event as much as possible. Participants jointly listened
to presentations, had live Q\&A, and attended other live events
associated with the conference. This was achieved by combining the
Zoom video conference software with the Slack communication
platform. Sessions took place on the dates originally planned for the
face-to-face event and we used one Zoom meeting for each of the
planned sessions, with a password provided to delegates. Fully
asynchronous, in contrast, could mean to put videos of the
presentations online to be watched by participants at the time that
fits them best and to have asynchronous Q\&A in Slack or a
similar tool. 

In retrospect, we are satisfied with choosing the synchronous
approach: interaction and discussion are key features of conferences
and although these cannot all be fully preserved online, we managed to
preserve them to a reasonable extent. Indeed, interesting discussions
emerged after many presentations and, to a lesser extent, also in
Slack. With more time to prepare and better tool support, we believe
that even more interaction can be fostered.


\paragraph{How do we deal with time differences?}
It seems difficult to deal with time differences when participants are
evenly distributed around the globe. In EDBT/ICDT, however, the bulk
of participants are from Europe, followed by North and South America, and Asia.  To
accommodate the relevant time zones, we opted for shorter days, about 5
to 6.5 hours, rather than the expected length (8-9 hours) in a face-to-face meeting.  The days were centered
around noon CET, which allowed attendees from other continents to
attend most sessions without major hassles. For some talks, we made
adjustments to the schedule in order to let the speaker present live,
e.g.\ from North America.  For example, keynote talks took place at
different times on different days as keynote speakers were from
different time zones.

In the future, one could try to adapt the program even more carefully
to speakers' time zones. In a Eurocentric conference such as EDBT/ICDT,
centering around noon CET is a natural thing to do, but speakers from
remote time zones could still not attend early and late sessions. This
could be alleviated by having even shorter conference days, at the
expense of stretching the conference over a longer time period.


\paragraph{How long should sessions and presentations be?}
Research sessions generally spanned an hour with the net talk length
for each papers being 10 and 12 minutes for EDBT and ICDT,
respectively. In comparison, EDBT and ICDT allot 20 and 25 minutes for
presentations in face-to-face conferences. The main reason for having
shorter talks was that we expected longer online sessions to be very
tiring for attendees. Another reason is that shorter talks help in
achieving shorter conference days to deal with time zone issues as
discussed above.

In retrospect, we were happy with the shorter presentations and the
sessions didn’t seem too long. This impression was confirmed by
the participants in our survey (Section~\ref{sec:survey}).

\paragraph{Should talks be recorded or given live?}
Following our guiding principle \eqref{GP}, we opted for pre-recorded
talks that we then streamed live from a central place with a high
capacity internet connection. Our aim was to minimize the probability
of technical problems that might result from participants having
differing internet connection quality and not being sufficiently
familiar with the Zoom software. We suggested to use Zoom to record
videos, with the speaker visible, which also helped to get
participants acquainted with Zoom.  We considered using Microsoft CMT
for video upload, but ended up using a simple sFTP solution as CMT has
a file size limit of 100MB. We checked the quality of the videos
beforehand. We plan to make the videos publicly available after the
conference.

Mostly this worked well, with many good quality videos being
submitted. On the one hand, pre-recording talks seemed to result in
presentations that were well planned, to the point, and with almost no
slips of the tongue. On the other hand, the talks were sometimes more
monotonous and less dynamic. Some presentations, including 2 keynotes
and 1 tutorial were presented live.
The keynotes and the tutorial went
fluently, but in one of the workshops there were some technical issues
with people presenting live. Practice sessions seem necessary for live
presentations.

\paragraph{How can questions be managed?}

Zoom has two modes, the \emph{meeting} mode and the \emph{webinar} mode.  In the
beginning of the conference, we used the webinar mode. This mode has a
text-based Q\&A facility that allows participants to type their
questions and to upvote questions asked by other participants. We then
had the people with the most popular questions ask them
face-to-face. Some participants asked questions in the Zoom chat, but this
rendered the chat (which sometimes uses pop-ups) distracting during
presentations. Especially when talks are very short, the smallest distraction
can bring listeners off-track. We
also generated one Slack channel per session, where question and
discussion could continue ``offline''.  Slack was a welcome
technological addition, which speakers also used to post their slides
after the talk.

The Q\&A facility in Zoom is quite good in principle, but it is only
available in webinar mode. For smaller, parallel sessions we preferred
a more informal approach based on Zoom's meeting mode, which we
switched to on the second day. In that mode, participants can see a
list of the names of the other participants and they can activate
their sound and video (but there is still a meeting host who can mute
everyone, e.g.\ when a talk starts). In meeting mode, we simply asked
participants to switch on their camera and raise their hand to
indicate that they want to ask a question, as in a real conference. We
encouraged people to also switch on their camera after each talk even
if they did not want to ask a question, the aim being to create a
community feeling, which was quite successful. Many people also
switched on their cameras at the beginning and end of each
sessions. We still used webinar mode for larger audiences, such as the
keynotes.

Retrospectively, we strongly prefer meeting mode and an informal
approach to Q\&A whenever the audience is of moderate size, say up to
50 participants. This brought much more interaction. Interestingly,
quite intense discussions emerged after some talks, probably even more
intense than in a face-to-face meeting. This might be due to the group
feeling created by Zoom meeting mode when several people have switched
on their camera, whereas in a face-to-face event, the few people who
are interested in an in-depth discussion of a presentation might sit
far apart from each other, with much less of a group feeling.


\paragraph{How can sessions be chaired?}
Following our guiding principle \eqref{GP}, we started with only a
Zoom host and no session chair. It turned out, however, that the
single host is rather busy with running the session, playing the
videos, and monitoring the chat. As the event went by, the technology
was holding up, and internet connections seemed to be sufficiently
stable, session chairs were introduced to manage discussions while a
separate Zoom host was technically managing the Zoom session, and this
was felt to be successful.

\paragraph{Can there be a social programme?}
The programme only had very short coffee breaks of 15 minutes, to make
the conference days shorter. There was no joint activity during the
coffee breaks apart from using the Slack channels.  We held two “Bring your
own beer” receptions, one on the opening evening of ICDT, and one on
the opening evening of EDBT, where evening refers to the CET time
zone. In the receptions, people arrived in an online session and were
assigned at random to Zoom breakout groups, to allow smaller group
interactions.

Retrospectively, we would go for longer coffee breaks to make the
conference days less exhausting. The receptions seemed to work well,
given the circumstances, and this was also confirmed in our
survey (Section~\ref{sec:survey}). There might well be scope for having more sessions with
opportunities for extended, informal, interactions. It might be
interesting to use other, technically more sophisticated tools for
this, such as Online Town.

\paragraph{How should short/poster papers be handled?}
We decided to waive short advert videos, and use Slack for asynchronous
discussions. We didn’t have a clear idea how to run more interactive sessions
that would simulate a poster session, so the short paper session turned into a
collection of short videos (26 in all) back-to-back with no intermediate Q\&A.

The session was well attended, with over 50 people there throughout. However, it
was hard work to sit through so many, diverse, videos, and the Slack channel was
not especially busy. Retrospectively, we would be tempted to simulate a poster
session in a more realistic way to enable deeper interactions. One way of doing
this could be to have each poster participant create their own Zoom meeting from
within Slack such that participants can use Slack to easily switch between the
rooms.

\paragraph{What should be the approach to demos?}
We adopted the conference model of videos of the demo in 15 minute slots. Each
demo was a 10 minute video, with 5 minutes for questions. In retrospect, the
videos were alright, but there wasn’t the chance for extended discussions that
are associated with demo sessions. Given that technology and internet
connections were more stable than we expected, we would be tempted to try having
each demo participant create their own Zoom meeting from within Slack such that
participants can use Slack to easily switch between the rooms. Again, this more
closely reflects the experience at a face-to-face demo session.

\paragraph{What should be the approach to keynotes and tutorials?}
Keynotes and tutorials were in 1-hour slots, some of them with gaps every 15 minutes for questions. A complete hour of presenting seemed rather long. Furthermore, in an online webinar everyone feels close to the presenter, so some people are more inclined to ask questions. As such, one may want to plan more discussion time. 

Two keynotes used videos, two were presented live. Concerning tutorials, three of them used pre-recorded videos, one was presented live. The division of talks into parts was felt to have been a success for both keynotes and tutorials in order to let people chime in with questions. 

\paragraph{What do we do if the meeting host has technical problems?}
Before the conference, we tested how the software platform reacts if the meeting
host drops out (e.g. by losing the internet connection). In our case, we
observed that the meeting can still continue, but the host’s video freezes. In
order to avoid major technical problems, we reached out to back-up hosts for
each session, who received a crash course on how to handle the software platform
about one week before the conference. We also wrote a general guide for session
hosts on how to set up all the parameters to make the sessions work the way we
want it. This technical aspect of running a conference certainly requires some
practice and we highly recommend thorough preparation.

\newpage
\section{Attendee Feedback Survey}\label{sec:survey}
We ran a feedback survey after the conference, which was answered by 114
participants (over 42\% of the registered participants). Here, we present a
selection of its results.

\medskip
\textbf{What is your current occupation? (114 responses)}
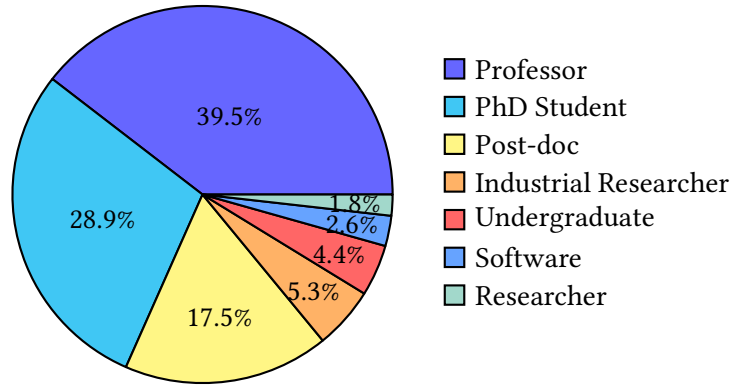
\begin{center}
\begin{tikzpicture}
\pie[text=legend, radius=2.5] {
  39.5/ Professor,
  28.9/ PhD Student,
  17.5/ Post-doc,
  5.3/ Industrial Researcher,
  4.4/ Undergraduate / Master's student,
  2.6/ Software / Hardware developer,
  1.8/ Researcher
  }
\end{tikzpicture}
\end{center}

\medskip
\textbf{Do you have a PhD? (114 responses)}
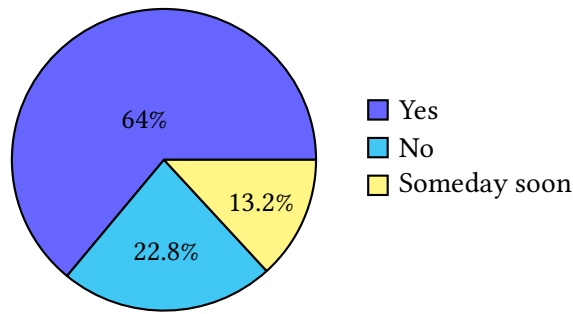
\begin{center}
\begin{tikzpicture}
\pie[text=legend, radius=2] {
  64/ Yes,
  22.8/ No,
  13.2/ Someday soon
  }
\end{tikzpicture}
\end{center}

\medskip
\textbf{Which continent were you on during the conference? (114 responses)}
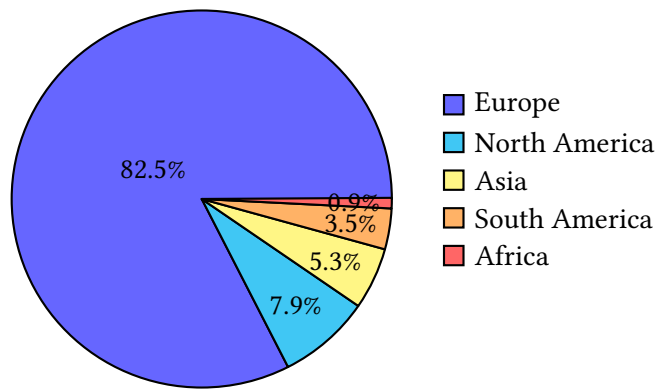
\begin{center}
\begin{tikzpicture}
\pie[text=legend, radius=2.5] {
  82.5/ Europe,
  7.9/ North America,
  5.3/ Asia,
  3.5/ South America,
  0.9/ Africa
  }
\end{tikzpicture}
\end{center}

\textbf{Which plenary sessions did you attend? (98 responses)}

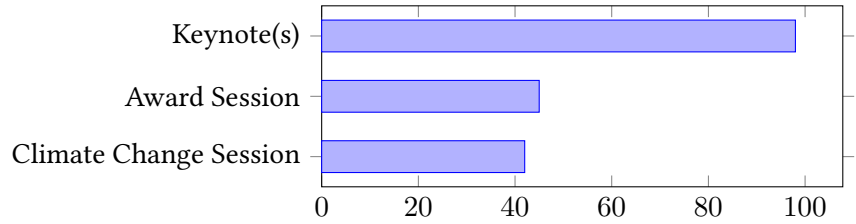
\pgfplotstableread[row sep=\\,col sep=&]{
  session & attendants  \\
  Keynote(s)  & 98 \\
  Award Session & 45 \\
  Climate Change Session & 42 \\
}\sessiondata

\begin{center}
  \begin{tikzpicture}
    \begin{axis}[
      xbar,
      symbolic y coords={Climate Change Session,Award Session,Keynote(s)},
      ytick=data,
      bar width=12pt,
      enlarge y limits={abs=4mm},
      y=8mm,
      xmin=0
      ]
      \addplot table[y=session,x=attendants]{\sessiondata};
    \end{axis}
  \end{tikzpicture}
\end{center}

\medskip
\textbf{How many ICDT-only sessions did you attend? (114 responses)}

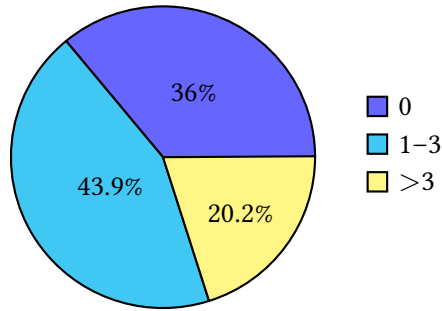
\begin{center}
  \begin{tikzpicture}
    \pie[text=legend, radius=2] {
      36/ 0,
      43.9/ 1--3,
      20.2/ $>$3
    }
  \end{tikzpicture}
\end{center}

\medskip
\textbf{How many EDBT-only sessions did you attend? (114 responses)}

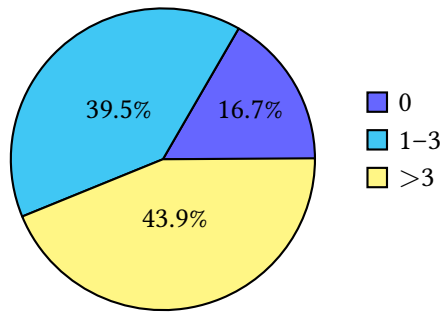
\begin{center}  
  \begin{tikzpicture}
    \pie[text=legend, radius=2] {
      16.7/ 0,
      39.5/ 1--3,
      43.9/ $>$3
    }
  \end{tikzpicture}
\end{center}

\medskip \noindent \textbf{Would you have attended more sessions if the conference would
  have been physical? (114 responses)}

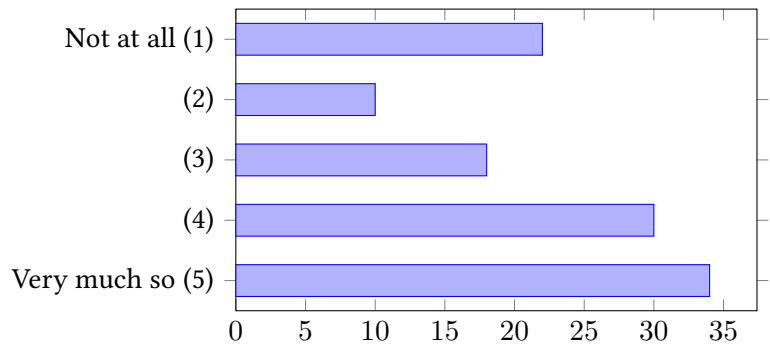
\pgfplotstableread[row sep=\\,col sep=&]{
  choice & data  \\
  Not at all (1) & 22 \\
  (2) & 10 \\
  (3) & 18 \\
  (4) & 30 \\
  Very much so (5) & 34\\
}\moresessiondata

\begin{center}
  \begin{tikzpicture}
    \begin{axis}[
      xbar,
      symbolic y coords={Very much so (5), (4), (3), (2), Not at all (1)},
      ytick=data,
      bar width=12pt,
      enlarge y limits={abs=4mm},
      y=8mm,
      xmin=0
      ]
      \addplot table[y=choice,x=data]{\moresessiondata};
    \end{axis}
  \end{tikzpicture}
\end{center}

\newpage

\noindent \textbf{How did the online video presentations compare to conventional
  conference talks? (114 responses)}

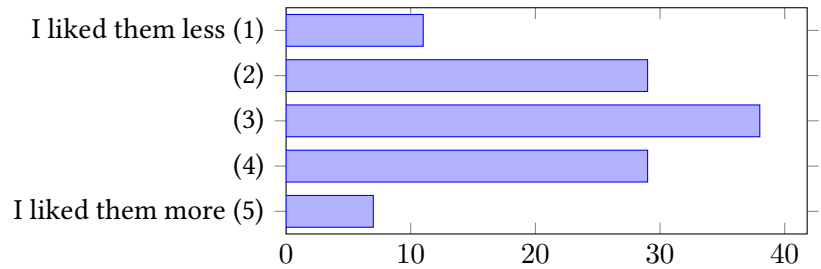
\pgfplotstableread[row sep=\\,col sep=&]{
  choice & data  \\
  I liked them less (1) & 11 \\
  (2) & 29 \\
  (3) & 38 \\
  (4) & 29 \\
  I liked them more (5) & 7\\
}\talklikedata

\begin{center}
  \begin{tikzpicture}
    \begin{axis}[
      xbar,
      symbolic y coords={I liked them more (5), (4), (3), (2), I liked them less (1)},
      ytick=data,
      bar width=12pt,
      enlarge y limits={abs=3mm},
      y=6mm,
      xmin=0
      ]
      \addplot table[y=choice,x=data]{\talklikedata};
    \end{axis}
  \end{tikzpicture}
\end{center}

Notice that this is a rather even binomial distribution, slightly weighted to
the negative.

\bigskip
\noindent
\textbf{What is the ideal length of a research talk for an
  online conference? (111 responses)}

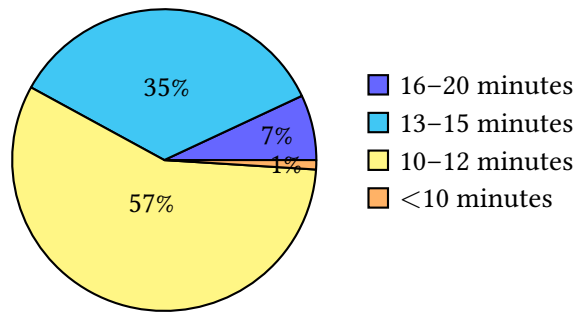
\begin{center}  
  \begin{tikzpicture}
    \pie[text=legend, radius=2] {
      7/ 16--20 minutes,
      35/ 13--15 minutes,
      57/ 10--12 minutes,
      1/ $<$10 minutes
    }
  \end{tikzpicture}
\end{center}

\noindent \textbf{As an attendant, which kind of model(s) do you prefer? Multiple choices possible.\\
(113 responses)}

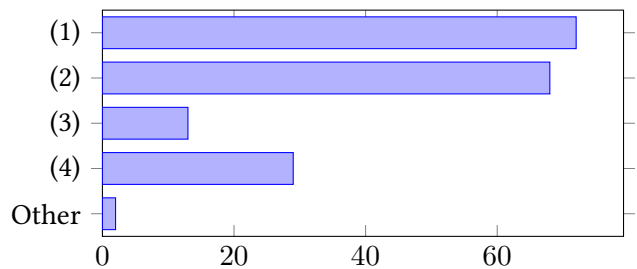
\pgfplotstableread[row sep=\\,col sep=&]{
  choice & data  \\
  (1)  & 72 \\
  (2) & 68 \\
  (3) & 13 \\
  (4) & 29 \\
  Other & 2\\
}\desiredmodel

\begin{center}
  \begin{tikzpicture}
    \begin{axis}[
      xbar,
      symbolic y coords={Other,(4),(3),(2),(1)},
      ytick=data,
      bar width=12pt,
      enlarge y limits={abs=3mm},
      y=6mm,
      xmin=0 
      ]
      \addplot table[y=choice,x=data]{\desiredmodel};
    \end{axis}
  \end{tikzpicture}
\end{center}
\begin{enumerate}[(1)]
\item Live talks, live Q\&A, and having recorded talks available after the
  conference
\item Streamed talks, live Q\&A, and having the videos available after the
  conference
\item Talk videos available beforehand and asynchronous Q\&A (e.g.\ over Slack)
\item  Talk videos available beforehand and live Q\&A in discussion sessions
\end{enumerate}

\newpage

\noindent
\textbf{Which meeting mode do you prefer for keynotes? (111 responses)}

\medskip
\noindent  
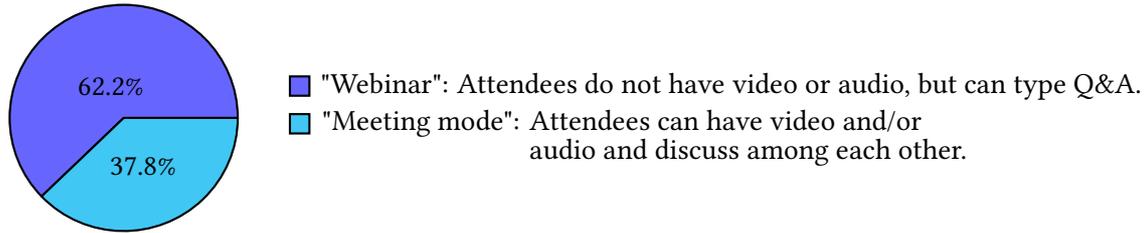
\begin{tikzpicture}
    \pie[text=legend, radius=1.5] {
      62.2/ {"Webinar": Attendees do not have video or audio, but can type Q\&A.},
      37.8/ 
    }
    \node at (7,-.25) {
      "Meeting mode": \hspace{-4mm}
      \adjustbox{valign=t}{
        \begin{tabular}{l}
          Attendees can have video and/or\\[-1mm]
          audio and discuss among each other.
        \end{tabular}
      }
    };     
  \end{tikzpicture}

\noindent
\textbf{Which meeting mode do you prefer for research talks? (111 responses)}

\medskip
\noindent 
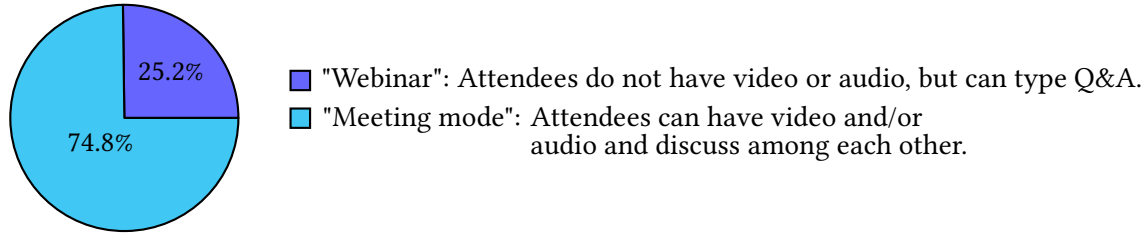
\begin{tikzpicture}
    \pie[text=legend, radius=1.5] {
     25.2/ {"Webinar": Attendees do not have video or audio, but can type Q\&A.},
     74.8/ 
   }
       \node at (7,-.2) {
      "Meeting mode": \hspace{-4mm}
      \adjustbox{valign=t}{
        \begin{tabular}{l}
          Attendees can have video and/or\\[-1mm]
          audio and discuss among each other.
        \end{tabular}
      }
    };     

  \end{tikzpicture}

\noindent
\textbf{Was the software infrastructure adequate for supporting EDBT/ICDT? (114 responses)}
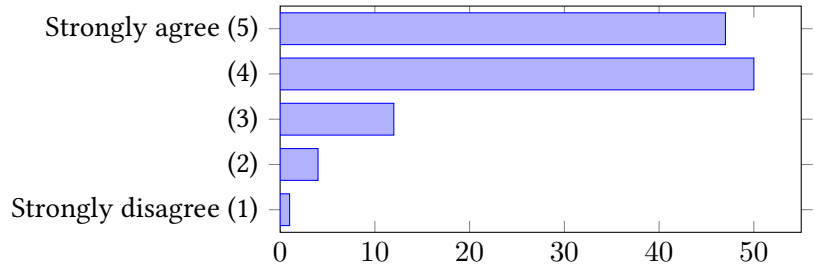
\pgfplotstableread[row sep=\\,col sep=&]{
  choice & data  \\
  Strongly disagree (1)   & 1 \\
  (2) & 4 \\
  (3) & 12 \\
  (4) & 50 \\
  Strongly agree (5) & 47\\
}\softwareadequate

\begin{center}
  \begin{tikzpicture}
    \begin{axis}[
      xbar,
      symbolic y coords={Strongly disagree (1),(2),(3),(4),Strongly agree (5)},
      ytick=data,
      bar width=12pt,
      enlarge y limits={abs=3mm},
      y=6mm,
      xmin=0 
      ]
      \addplot table[y=choice,x=data]{\softwareadequate};
    \end{axis}
  \end{tikzpicture}
\end{center}

\noindent\textbf{How many Slack channels have you joined during the whole conference?\\
  (109 responses)}
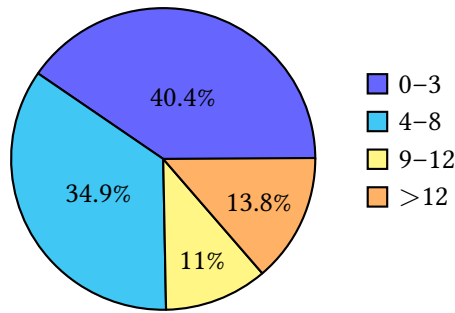
\begin{center}  
  \begin{tikzpicture}
    \pie[text=legend, radius=2] {
     40.4/ 0--3,
     34.9/ 4--8,
     11/ 9--12,
     13.8/ $>$12
    }
  \end{tikzpicture}
\end{center}

\newpage
\noindent
\textbf{Did the conference need more social interaction? (111 responses)}
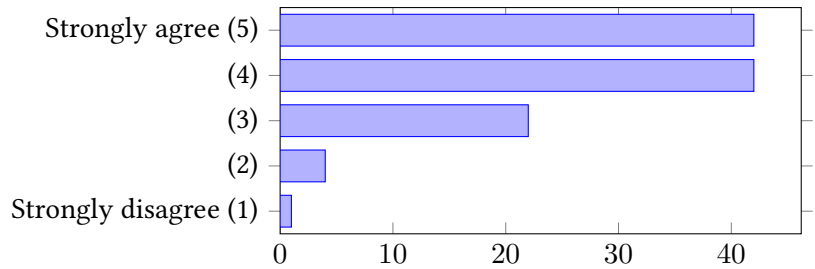
\pgfplotstableread[row sep=\\,col sep=&]{
  choice & data  \\
  Strongly disagree (1)   & 1 \\
  (2) & 4 \\
  (3) & 22 \\
  (4) & 42 \\
  Strongly agree (5) & 42\\
}\softwareadequate

\begin{center}
  \begin{tikzpicture}
    \begin{axis}[
      xbar,
      symbolic y coords={Strongly disagree (1),(2),(3),(4),Strongly agree (5)},
      ytick=data,
      bar width=12pt,
      enlarge y limits={abs=3mm},
      y=6mm,
      xmin=0
      ]
      \addplot table[y=choice,x=data]{\softwareadequate};
    \end{axis}
  \end{tikzpicture}
\end{center}

\noindent\textbf{Did you attend the Bring Your Own Beer sessions? (111 responses)}
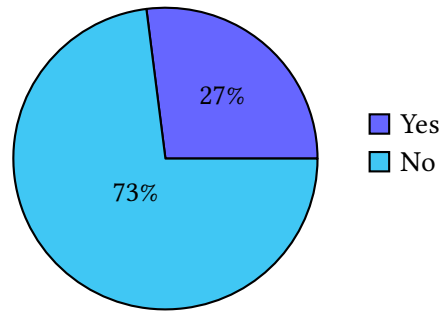
\begin{center}  
  \begin{tikzpicture}
    \pie[text=legend, radius=2] {
     27/ Yes,
     73/ No
    }
  \end{tikzpicture}
\end{center}

\noindent\textbf{Did you like the Bring Your Own Beer sessions? (30 responses)}
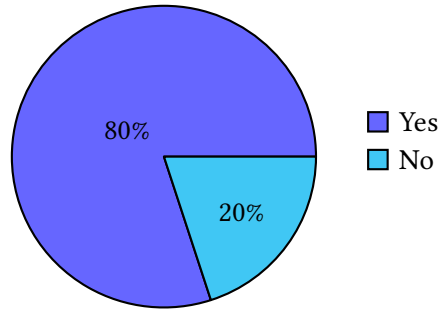
\begin{center}  
  \begin{tikzpicture}
    \pie[text=legend, radius=2] {
     80/ Yes,
     20/ No
    }
  \end{tikzpicture}
\end{center}
In this question, we only selected the answers from respondents who said that
they attended the Bring Your Own Beer sessions.

\medskip
\noindent\textbf{Do you have suggestions for more social interactions or
  networking?}

\noindent This was a free-text field in the questionnaire. We received suggestions such as:
\begin{itemize}
\item Breakout rooms during coffee breaks, where you can see who is in which room.
\item Dagstuhl-like breakout rooms (randomly put people together).
\item Focused discussion rooms, e.g., per scientific area or per need (``looking
  for a phd student'')
\item Dedicated Ask Me Anything (AMA) with speakers (at least for keynotes).
\item Playing some online game together.
\item Imitate a "social event" like watching a live-streamed video/concert
  together.
\end{itemize}
Given the short timespan, our planning of EDBT/ICDT 2020 indeed focused mostly
on simply making the scientific part of the conference work. There is much more
that can be done in terms of social interaction in an online conference. Social interaction
sessions could act as a buffer alongside the programme, i.e., they could be
running before / after / during the scientific programme to further accommodate
time zone differences.

\medskip
\noindent\textbf{As a presenter of a research talk, would you prefer to present live or
  with a pre-recorded talk? (80 responses)}
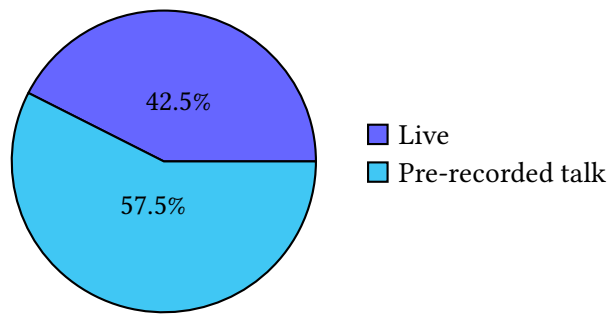
\begin{center}  
  \begin{tikzpicture}
    \pie[text=legend, radius=2] {
     42.5/ Live,
     57.5/ Pre-recorded talk
    }
  \end{tikzpicture}
\end{center}

\noindent\textbf{As a presenter of a demo, would you prefer to present live or
  with a pre-recorded talk? (36 responses)}
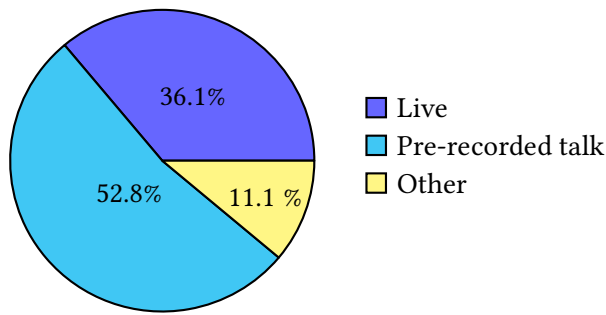
\begin{center}  
  \begin{tikzpicture}
    \pie[text=legend, radius=2] {
     36.1/ Live,
     52.8/ Pre-recorded talk,
     11.1 / Other
    }
  \end{tikzpicture}
\end{center}
Among the suggestions for ``Other'', respondents wrote suggestions like giving
live commentary on a pre-recorded presentation of the system, or having a short
(pre-recorded?) presentation to give the audience the context and then a live demo.

\newpage
\noindent\textbf{Assume that EDBT/ICDT would be held physically. Would you attend
  virtually if this option existed? (113 responses)}
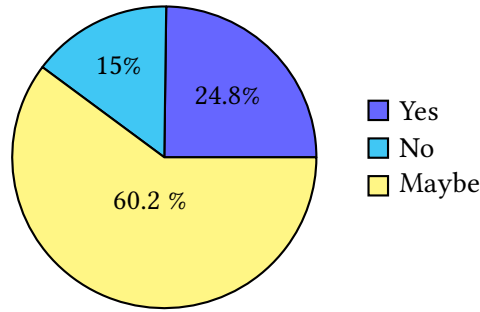
\begin{center}  
  \begin{tikzpicture}
    \pie[text=legend, radius=2] {
     24.8/ Yes,
     15/ No,
     60.2 / Maybe
    }
  \end{tikzpicture}
\end{center}

\noindent\textbf{For which reasons would you consider to attend a hybrid conference
  virtually?\\ (107 responses)}

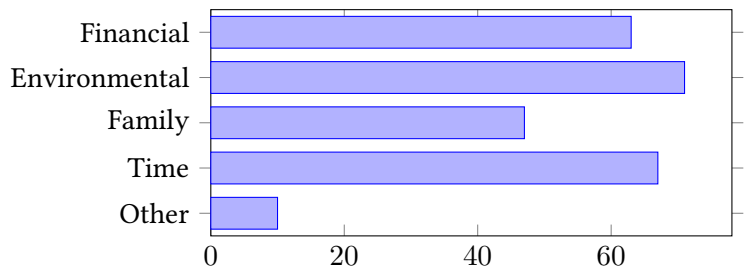
\pgfplotstableread[row sep=\\,col sep=&]{
  choice & data  \\
  Other & 10\\
  Time &  67\\
  Family & 47\\
  Environmental & 71\\
  Financial & 63 \\
}\hybridreasons

\begin{center}
  \begin{tikzpicture}
    \begin{axis}[
      xbar,
      symbolic y coords={Other, Time, Family, Environmental, Financial},
      ytick=data,
      bar width=12pt,
      enlarge y limits={abs=3mm},
      y=6mm,
      xmin=0
      ]
      \addplot table[y=choice,x=data]{\hybridreasons};
    \end{axis}
  \end{tikzpicture}
\end{center}

Among the other reasons were: being able to attend more conferences per year,
being able to join virtually even when you don't have a paper, being able to
network more, being able to buy virtual registrations for internship students,
no visa restrictions, less travelling.

\bigskip
\noindent\textbf{Assume that EDBT/ICDT would be held virtually only. Would you attend?}
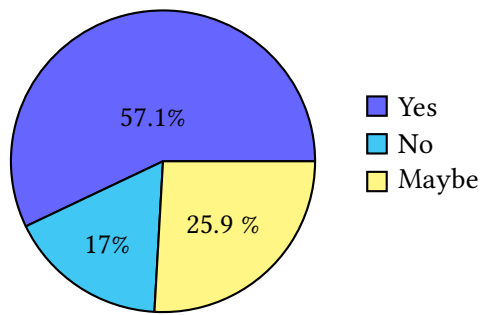
\begin{center}  
  \begin{tikzpicture}
    \pie[text=legend, radius=2] {
     57.1/ Yes,
     17/ No,
     25.9 / Maybe
    }
  \end{tikzpicture}
\end{center}

\newpage
\noindent\textbf{Would you support the idea of having hybrid conferences for reducing CO2
  emissions? (112 responses)}
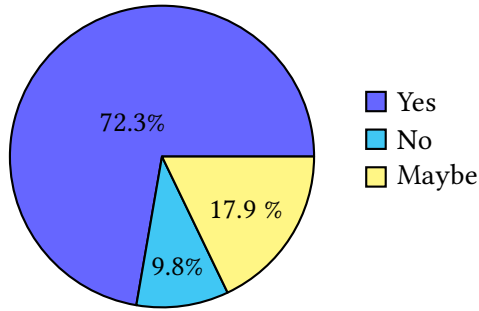
\begin{center}
  \begin{tikzpicture}
    \pie[text=legend, radius=2] {
     72.3/ Yes,
     9.8/ No,
     17.9 / Maybe
    }
  \end{tikzpicture}
\end{center}

\noindent\textbf{Would you support the idea of alternating physical meetings with purely
  virtual conferences for reducing CO2 emissions? (112 responses)}
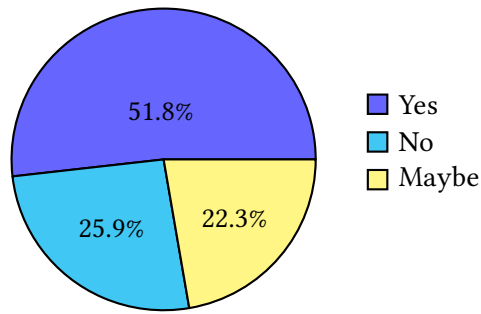
\begin{center}
  \begin{tikzpicture}
    \pie[text=legend, radius=2] {
     51.8/ Yes,
     25.9/ No,
     22.3/ Maybe
    }
  \end{tikzpicture}
\end{center}

\noindent\textbf{Was EDBT / ICDT 2020 better or worse than a typical other
  edition? (114 responses)}

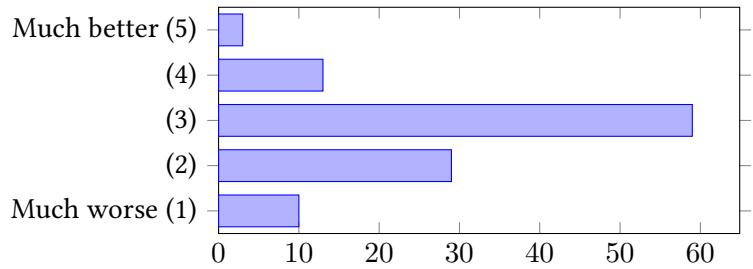
\pgfplotstableread[row sep=\\,col sep=&]{
  choice & data  \\
  Much worse (1)   & 10 \\
  (2) & 29 \\
  (3) & 59 \\
  (4) & 13 \\
  Much better (5) & 3\\
}\betterorworse

\begin{center}
  \begin{tikzpicture}
    \begin{axis}[
      xbar,
      symbolic y coords={Much worse (1),(2),(3),(4),Much better (5)},
      ytick=data,
      bar width=12pt,
      enlarge y limits={abs=3mm},
      y=6mm,
      xmin=0
      ]
      \addplot table[y=choice,x=data]{\betterorworse};
    \end{axis}
  \end{tikzpicture}
\end{center}

\noindent\textbf{Was attending EDBT/ICDT 2020 better or worse than what you expected a
  virtual conference to be like? (110 responses)}

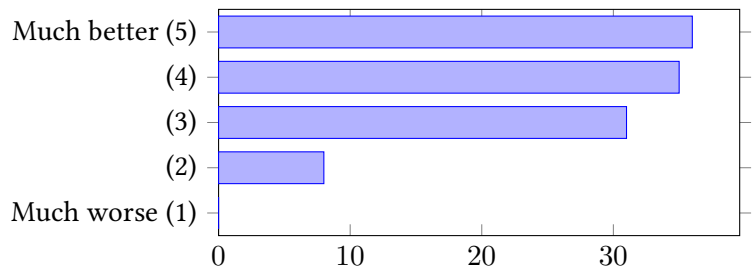
\pgfplotstableread[row sep=\\,col sep=&]{
  choice & data  \\
  Much worse (1)   & 0 \\
  (2) & 8 \\
  (3) & 31 \\
  (4) & 35 \\
  Much better (5) & 36\\
}\betterorworse

\begin{center}
  \begin{tikzpicture}
    \begin{axis}[
      xbar,
      symbolic y coords={Much worse (1),(2),(3),(4),Much better (5)},
      ytick=data,
      bar width=12pt,
      enlarge y limits={abs=3mm},
      y=6mm,
      xmin=0
      ]
      \addplot table[y=choice,x=data]{\betterorworse};
    \end{axis}
  \end{tikzpicture}
\end{center}

Judging from these last two questions, attendants found that the online
experience was indeed somewhat less than the physical experience, but at least
the experience was better than what was expected from a virtual conference.
Furthermore, we need to keep in mind that this edition was planned and organized
in just three weeks, without any external guidelines. There is room for
improvement.

\section{Conclusions and Advice for Future Events}


First-time organization of an online conference is a complicated matter,
especially under tight time constraints. In our case, a great effort of
coordination was needed and a task force (formed by the people co-authoring this
report) made the executive decisions and carried out the required work. On the
other hand, once the executive decisions have been made, the organizational
amount of work is reasonable. We therefore encourage other conferences to try out the
transition to an on-line mode in the short term. In the medium to long term,
on-line and/or hybrid conferences may help the community
reduce its CO2 footprint.

With a long-standing experience and
within a large time window, things can be arranged more carefully.
For instance, one could think about the following issues: 
\begin{itemize}
\item Carefully choosing the underlying technological platforms on which the conference has to be hosted. 
People are aware of security issues around Zoom but there is no available equivalent open-source tool that can host the same 
number of participants. Since a high number of participants is needed for plenary sessions, hopefully such open-source tools 
will be available in the long run. Dedicated platforms for scientific conferences are urgently needed in that respect. 

\item Several sessions that require tighter interactions, such as poster and demonstration sessions, need 
to be planned carefully. For instance, for demonstrations and posters, one would rely on the breakout rooms 
in Zoom to let people gather around a demo booth or a poster (with limited number of participants). If the poster or the demonstrated tool can 
be shared with the participants beforehand, the sessions can be also prepared 
in advance and be more fluent and interactive. Other sophisticated solutions, such as 
virtual reality and avatar-based video and chat tools, may be needed in the long
run. These tools would help reproducing the physical interactions needed for poster and demo sessions along with the 
serendipity of meeting people with similar interests at these sessions.

\item Networking would greatly benefit from having dedicated online sessions that are scheduled alongside the normal scientific sessions 
of the conference. Networking is truly the pitfall of an online event and this is especially deleterious for the junior members of our community. 
An idea would be to prepare networking well in advance and to pin interesting topics or discussions with colleagues of other universities and research teams (a sort of Pinterest specialized for 
scientific conferences). 

\end{itemize}

Finally, we are pleased to share our experience at online EDBT/ICDT 2020 and
eager to learn more about virtual scientific events in the near future. During
the climate change session, which has been hosted by the conference this year, we
had a lively and stimulating discussion about adopting CO2 plans for
conferences. One of the options there is to allow alternate virtual and
in-person events or hybrid (simultaneously virtual and in-person) events and
thus contribute to reducing the environmental footprint of scientific
conferences. Our on-line survey gives us two hopeful signs. First, there is a
significant support of the community for going on-line in order to reduce CO2
footprint, and second, attendees clearly found this year's conference better
than what they expected a virtual conference to be like.

Finally, in the spirit of moving open science and open access forward in the
right direction, the videos of the conference talks are going to be made
available to the general public. For EDBT, they can be linked directly from
the proceedings and ICDT is looking into a similar solution.



\bibliographystyle{ACM-Reference-Format}
\bibliography{references}

\end{document}